# A Graphene Quantum Dot with a Single Electron Transistor as Integrated Charge Sensor

*Ling-Jun Wang [§], Gang Cao [§], Tao Tu [§], Hai-Ou Li [§], Cheng Zhou [§], Xiao-Jie Hao [†], Zhan Su [§], Guang-Can Guo [§], Guo-Ping Guo [§,\*], and Hong-Wen Jiang [†,\*]*

§ Key Laboratory of Quantum Information, University of Science and Technology of China, Chinese Academy of Sciences, Hefei 230026, People's Republic of China

† Department of Physics and Astronomy, University of California at Los Angeles, California 90095, USA.

\* Corresponding authors. Email: gpguo@ustc.edu.cn, jiangh@physics.ucla.edu.

**ABSTRACT:** We have developed an etching process to fabricate a quantum dot and a nearby single electron transistor as a charge detector in a single layer graphene. The high charge sensitivity of the detector is used to probe Coulomb diamonds as well as excited spectrum in the dot, even in the regime where the current through the quantum dot is too small to be measured by conventional transport means. The graphene based quantum dot and integrated charge sensor serve as an essential building block to form a solid-state qubit in a nuclear-spin-free quantum world.

Graphene has attracted a lot of research interest because of its unique electronic properties which make it a promising candidate for future nanoelectronics.[1–3] Because of the weak spin-orbit coupling and largely eliminated hyperfine interaction in graphene, it is highly desirable to coherently control the spin degree of freedom in graphene nanostructures for spin-based quantum computation. Recently, there was a striking advance on experimental production of graphene single or double quantum dots,[4–7] which is an important first step towards such promise.

The measurement of individual electrons or its spins in GaAs quantum dots (QDs) have been realized by so-called charge detection via a nearby quantum point contact (QPC) or single electron transistor (SET).[8,9] In particular, the combination of high speed and very high charge sensitivity has made SET for use in studying a wide range of physical phenomena such as discrete electron transport,[8,10,11] qubit readout,[12,13] and nanomechanical oscillators.[14,15] So far most SETs have been using $Al/AlO_x/Al$ tunnel junctions, graphene SET reported here is technologically simple, reliable and can operate well above liquid-helium temperature, making it an attractive candidate for use in various charge detector applications.

Here we realize an all graphene nanocircuit integration with SET charge read-out for QDs. The QD and the SET in the same material are defined in a single etching step which enables optimized coupling and sensing ability. The SET is placed in close vicinity to the QD giving rise to a strong capacitive coupling between the two systems. Once an additional electron occupies the QD, the potential in the neighboring SET is modified by capacitive interaction that gives rise to a measurable conductance change. Even if charge transport through the QD is too small to be measured by conventional transport means, the SET charge sensor also allows measurements. These devices demonstrated here provide robust building blocks in a practical quantum information processor.

The graphene flakes were produced by mechanical cleaving of graphite crystallites by scotch tape and then were transferred to a highly doped Si substrate with a 100 nm $SiO_2$ top layer. Thin flakes were found by optical microscopy and single layer graphene flakes were selected by the Raman spectroscopy measurement. Next, a layer of poly(methyl methacrylate) (PMMA) is exposed by standard electron beam lithography (EBL) to form a designed pattern. The unprotected areas are carved by oxygen reactive ion etching. We used the standard EBL and lift off technique to make the ohmic contact (Ti/Au) on the present graphene devices. One of our defined sample structures with a quantum dot and proximity SET is shown in Figure 1. The quantum dot is an isolated central island of diameter 90 nm, connected by 30 nm wide tunneling barriers to source and drain contacts. Here, Si wafer was used as the back gate and there is also a side gate. The SET has a similar pattern while the conducting island has much larger diameter (180 nm). Electronic transport through both the devices exhibits Coulomb blockade characteristics with back/side gate voltage. The distance between CB peaks is determined by the sum of charging and quantum confinement energies and the former contribution becomes dominant for our devices with diameter > 100 nm. Accordingly, we refer to it as SET rather than QD. The device was first immersed into a liquid helium storage dewar at 4.2 K to test the functionality of the gates. The experiment was carried out in a top-loading dilution refrigerator equipped with filtered wiring and low-noise electronics at the base temperature of 10 mK. In the measurement, we employed the standard AC lock-in technique.

Figure 2a shows the conductance through the dot $G_{QD}$ for applied side gate voltage $V_{sg}$. Clear Coulomb blockade peaks are observed related to charging of the tunable dot on the graphene. The current through the dot becomes too small to be measurable for side gate voltages in the range of 0.2-0.5 V. Figure 2b shows the conductance through the SET versus side gate voltage $V_{sg}$. The SET is close as possible to the QD and in this way charging signals of the dot were detected by tracking the change in the SET current. The addition of one electron to the QD leads to a pronounced change of the conductance of the charge detector by typically 30%. The slope of the SET conductance is the steepest at both sides of its Coulomb blockade resonances giving the best charge read-out signal. To offset the large current background, we used a lock-in detection method developed earlier for GaAs dot.[16] A square shaped pulse was superimposed on the dc bias on side gate voltage $V_{sg}$. A lock-in detector in sync with the pulse frequency measured the change of SET current due to the pulse modulation. Figure 2C shows a typical trace of the lock-in signal of the transconductance through the SET $dI_{SET}/dV_{sg}$. These sharp spikes or dips originate from the change of the charge on the dot by one electron. It shows essentially the same features as Figure 2a but is much richer especially in the regime, where the direct dot current is too small to be measured. We also note that the individual charge events measurement has been demonstrated in graphene QD with QPC detector based on graphene nanoribbon.[17]

More quantitative information on the system can be obtained from the measurement of the height response of one certain peak in Figure 2c as a function of the modulating pulse frequency on the side gate. The resulting diagram for the SET $dI_{SET}/dV_{sg}$ gain magnitude is shown in Figure 3. The green dashed line indicates the gain of 0.707 (-3 dB), corresponding approximately to the bandwidth of 600 Hz of the SET device. By applying a signal of $5 \times 10^{-2}$ electrons on the back gate of SET and measuring the signal with a signal-to-noise ratio of 1, we achieved a charge sensitivity of $10^{-3} e/\sqrt{Hz}$ which is a nice match to that obtained previously in a GaAs QD and superconducting Al SET detector system.[18]

The information contained in the signal goes beyond simple charge counting. For instance, the stability diagram measurement can reveal excited states, which is crucial to get information of the spin state of electrons on a quantum dot.[19] Figure 4a shows Coulomb diamonds for the conductance $G_{QD}$ through the dot versus bias voltage $V_{sd}$ and side gate voltage $V_{sg}$. For comparison, Figure 4b shows the transconductance of the SET $dI_{SET}/dV_{sg}$ as a function of the same parameters. A perfect match between the QD transport measurements and the detector signal is observed. Moreover, the discrete energy spectrum of the graphene quantum dot is revealed by the presence of additional lines parallel to the diamond edges. These lines indicate the quantum dot is in the high bias regime where the source-drain bias is so high that multiple dot levels involving the excited states can participate in electron tunneling.[20] The excited states become much more visible in the charge detector signal than the direct measurement. All of these features have been seen in GaAs QD with QPC[20] but here we achieve the goal with an all graphene nanocircuit of QD with SET. In the previous case, the QD and QPC detector are separated by typically 100 nm in width. In the present case, the SET detector is 50 nm from the edge of the QD. Therefore it is expected that the capacitance coupling between QD and SET is enhanced compared to the conventional case realized in semiconductor QD and QPC. This enhanced coupling leads to a larger signal-to-noise ratio of the SET detector signal that can be exploited for time resolved charge measurement or charge/spin qubit readout on the QD.

In summary, we have presented a simple fabrication process that facilitates a quantum dot and highly sensitive single electron transistor charge detector with the same material of graphene. Typically the addition of a single electron in QD would result in a change in the SET conductance of about 30%. This way the charging events measured by the charge detector and direct transport through the dot merge perfectly and more information beyond the conventional transport means is also guaranteed. The devices demonstrated here represent a fascinating avenue to realize a more complex and highly controllable electronic nanostructure formed from molecule conductors such as graphene.

**Acknowledgements:** This work was supported by the National Science Foundation (DMR-0804794), the National Basic Research Program of China (Grants No. 2009CB929600, No. 2006CB921900), and the National Natural Science Foundation of China (Grants No. 10804104, No. 10874163, No. 10934006).

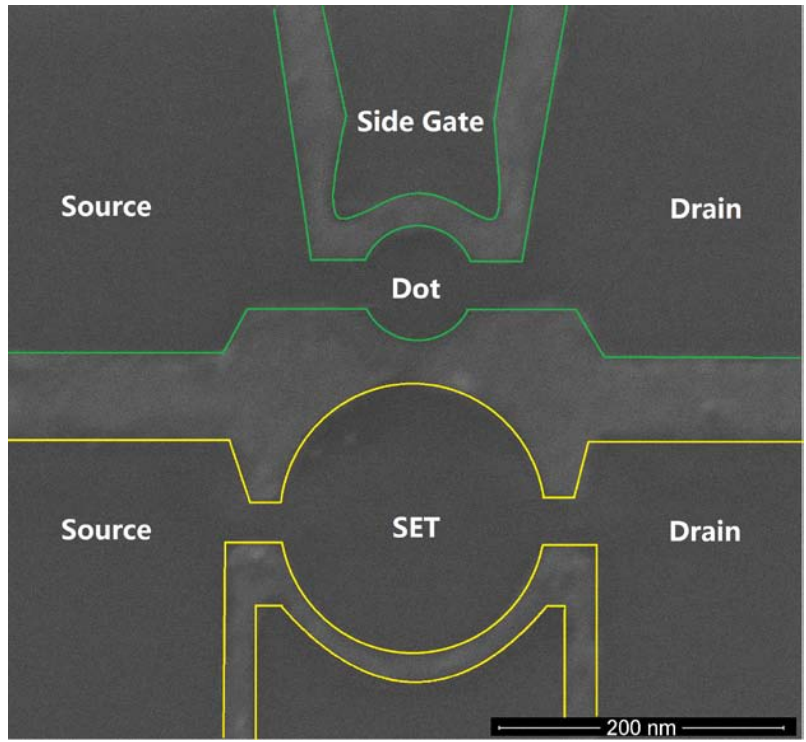

(a)

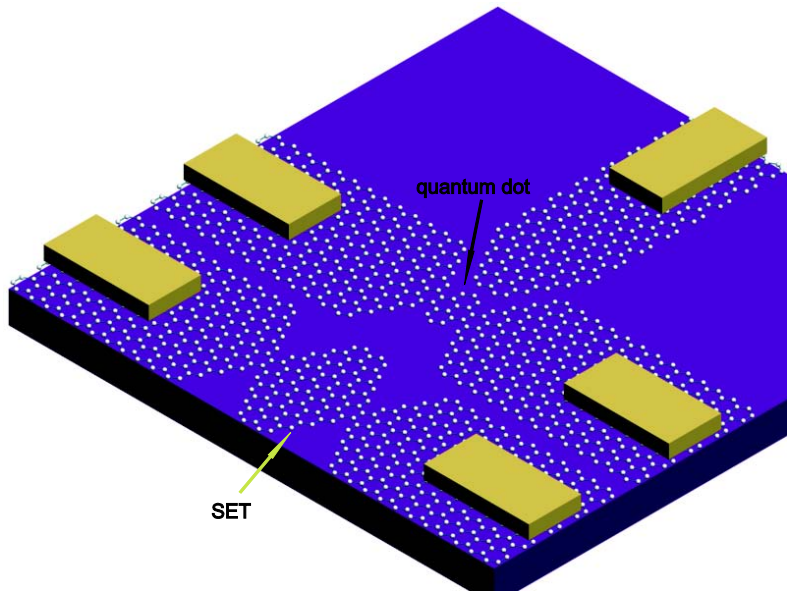

(b)

**FIGURE 1.** (a) Scanning electron microscope image of the etched sample structure. The bar has a length of 200 nm. The upper quantum dot as main device has a diameter of 90 nm while the bottom single electron transistor as charge sensor has a diameter of 180 nm. The bright lines define barriers and the side gates. (b) Schematic of a representative device.

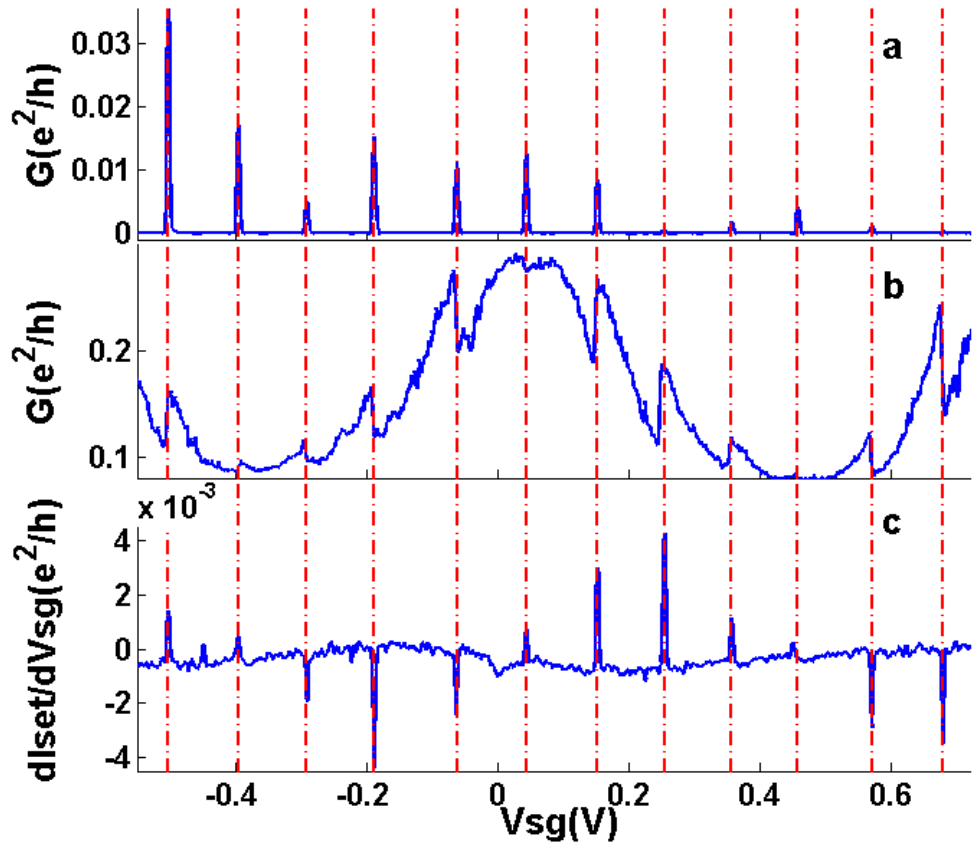

**FIGURE 2.** (a) Conductance through the quantum dot vs the side gate voltage. (b) The example of conductance through the single electron transistor for the same parameter ranges as in panel a. The steps in conductance are well aligned with the Coulomb blockades in panel a and are about 30% of the total signal. (c) Transconductance of the single electron transistor for the same parameter ranges as in panel a. The spikes and dips indicate the transitions in the charge states by addition of single electron in quantum dot. In particular, the charge detection can allow measurement in the regime (0.2-0.5 V) where the current through the dot is too small to be measurable by direct means. Data in panels a, b and c were recorded simultaneously during a single sweep. The vertical red dashed lines are a guide for the eyes to relate features in these graphs.

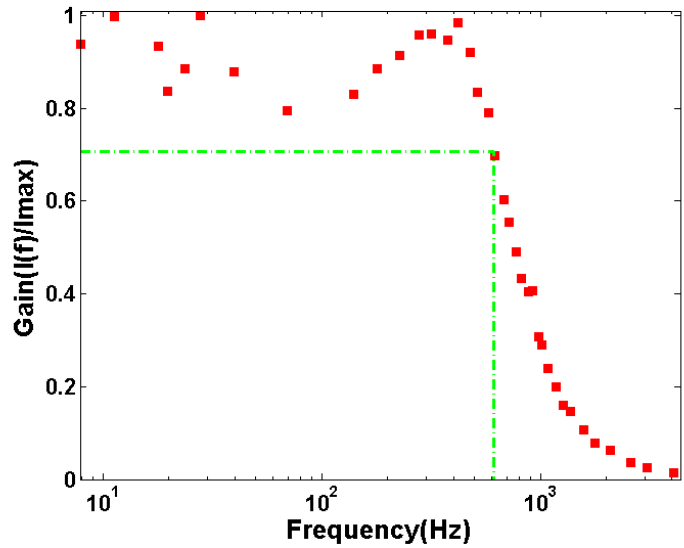

**FIGURE 3.** The SET $dI_{SET}/dV_{sg}$ gain magnitude with the modulating pulse frequency. The green dashed line illustrates the bandwidth of the SET device about 600 Hz corresponding to a gain of 0.707(-3 dB). Due to the stray capacitances, the response decreases rapidly after 600 Hz. The frequency axis is logarithmically scaled for clarity.

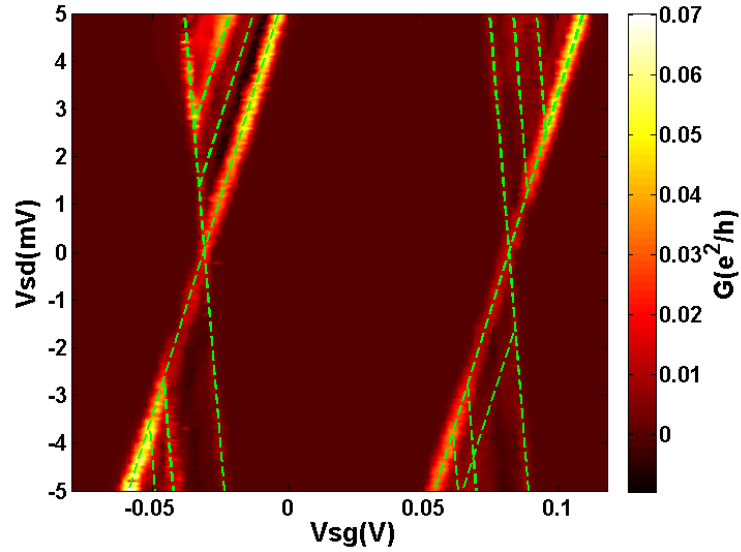

(a)

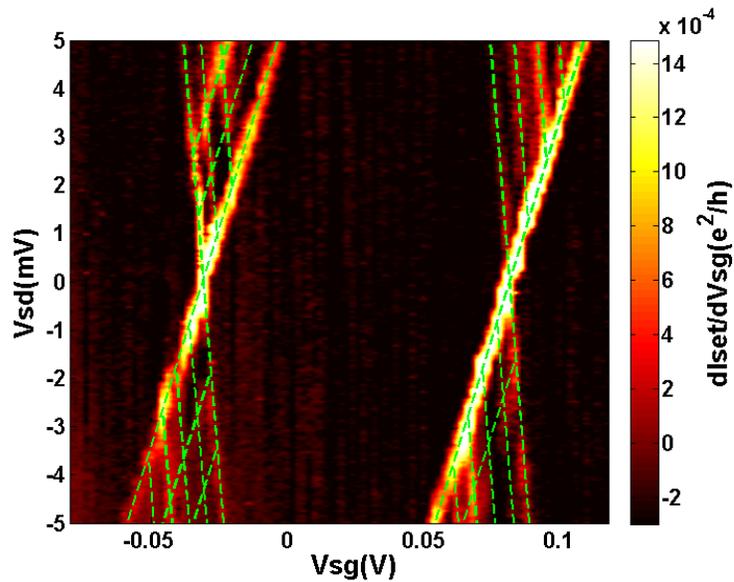

(b)

**FIGURE 4.** (a) Plot of the different conductance of the quantum dot vs bias and gate voltage applied on the dot. From the lines parallel to the edges of Coulomb diamonds, we can identify the excited states. (b) Transconductance of the single electron transistor for the same parameters as in panel a. Perfect matching with panel a and more excited spectrum resolved in the signal indicate the single electron transistor can be used as a highly sensitive charge detector. Data in panels a and b were recorded simultaneously during a single sweep. Green dashed lines are a guide for identifying the excited states.